\documentclass[conference]{IEEEtran}
\usepackage[utf8]{inputenc}
\usepackage[T1]{fontenc}
\usepackage{graphicx}
\usepackage{amsmath}
\usepackage{amssymb}
\usepackage{hyperref}
\usepackage{booktabs}
\usepackage{xcolor}
\usepackage{listings}

\hypersetup{
    colorlinks=true,
    linkcolor=blue,
    citecolor=blue,
    urlcolor=blue
}

\title{From Reactive to Autonomous: Evolution of AI Operations\\in Cloud Network Infrastructure}

\author{
\IEEEauthorblockN{Arun Malik}
\IEEEauthorblockA{
Microsoft Azure Networking\\
Email: arunma@microsoft.com\\
ORCID: 0009-0005-6650-6711
}
}

\begin{document}

\maketitle

\begin{abstract}
The operational model for cloud network infrastructure has undergone a fundamental transformation over the past decade. What began as manual, human-driven troubleshooting has evolved through scripted automation, rule-based systems, and AI-assisted operations into fully autonomous incident resolution. This paper traces the evolution of AI operations (AIOps) in cloud network infrastructure, identifying the architectural patterns, organizational challenges, and technical inflection points that enabled each generational transition. Drawing from production experience operating network infrastructure at hyperscale, we present a maturity model that characterizes five distinct operational generations, analyze the technical and organizational barriers that impede transitions between generations, and document the metrics that indicate readiness for increased autonomy. We show that the path from reactive to autonomous operations is not merely a technology problem but requires co-evolution of tooling, trust frameworks, knowledge management practices, and operational culture. Our findings provide a practical roadmap for infrastructure organizations seeking to adopt progressively autonomous AI operations.
\end{abstract}


\section{Introduction}

Cloud network infrastructure has grown to a scale that challenges every assumption of traditional operations models. A major cloud provider today operates tens of millions of network devices across hundreds of data centers, generating millions of telemetry signals per second and thousands of operational incidents per day. The mismatch between the growth rate of infrastructure and the growth rate of operational expertise has created a persistent and widening gap that no amount of hiring can close.

The industry response to this gap has been a progressive adoption of AI and automation technologies, often grouped under the banner of AIOps (Artificial Intelligence for IT Operations). However, the term AIOps obscures significant variation in maturity, capability, and operational authority. A system that generates alert summaries and a system that autonomously remediates hardware failures are both called AIOps, yet they represent fundamentally different operational paradigms with different architectural requirements, trust models, and failure modes.

\begin{figure*}[t]
\centering
\includegraphics[width=0.85\textwidth]{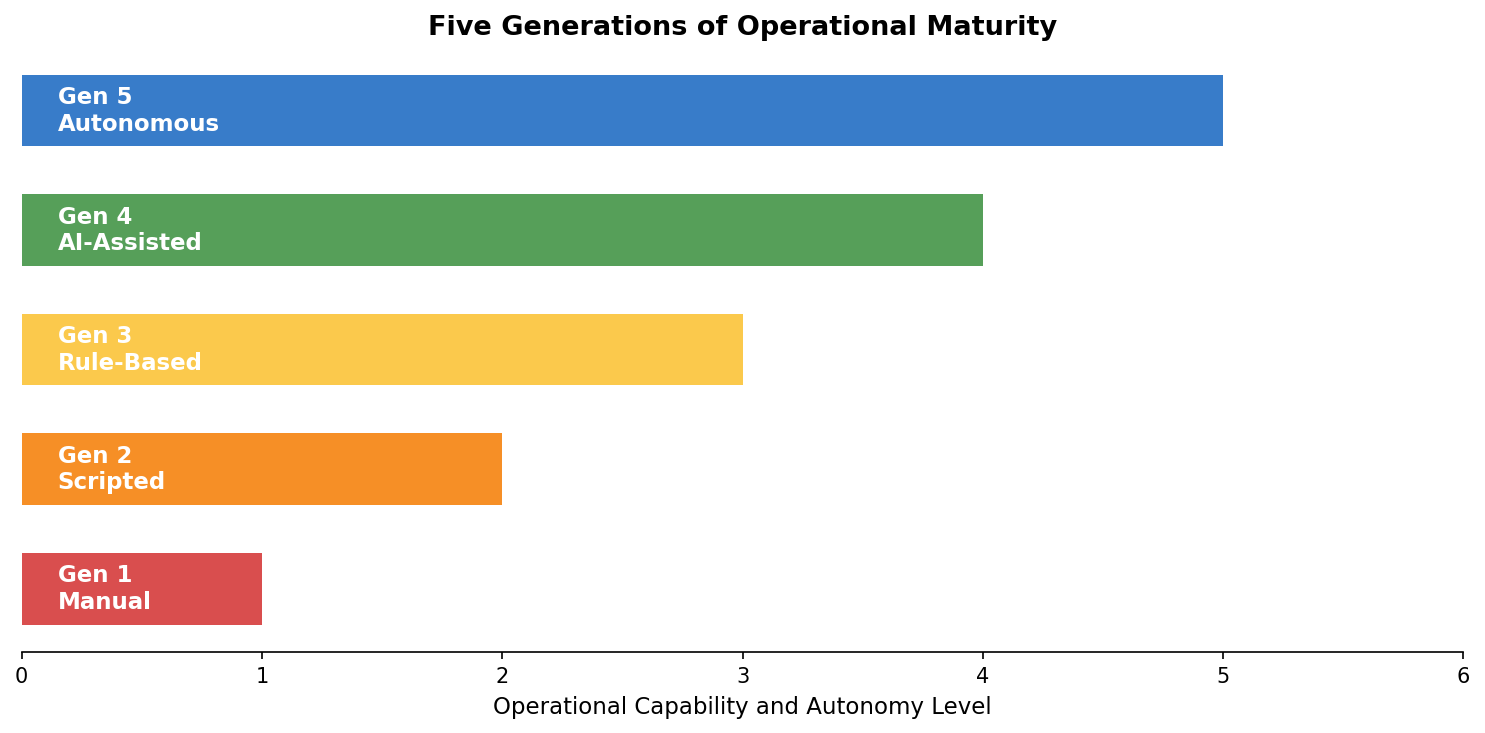}
\caption{Evolution of operational maturity from manual to autonomous operations. Each generation represents an architectural discontinuity with distinct human-AI authority models and characteristic resolution times.}
\label{fig:evolution}
\end{figure*}

This paper provides a structured framework for understanding the evolution from reactive, human-driven operations to autonomous, AI-driven operations in cloud network infrastructure. We identify five distinct operational generations (Table~\ref{tab:generations}, Figure~\ref{fig:evolution}), analyze the technical and organizational transitions between them, and document lessons learned from navigating this evolution in production at hyperscale.

\subsection{Motivation}

Despite significant industry investment in AIOps, most organizations remain stuck at early maturity levels. Surveys consistently report that fewer than 15\% of enterprises have achieved meaningful autonomous operations~\cite{gartner2024aiops}. The barriers are not purely technical. Organizations face challenges in building trust in AI systems, encoding operational knowledge in machine-executable forms, designing safety mechanisms that enable rather than prevent autonomy, and evolving operational culture to accommodate AI agents as first-class participants.

By documenting the full evolutionary arc and the specific transitions between generations, we aim to provide organizations with a practical roadmap that addresses both technical architecture and organizational readiness.

\subsection{Contributions}

This paper makes the following contributions:

\begin{enumerate}
    \item A five-generation maturity model for AI operations that characterizes distinct architectural patterns, capability levels, and operational authority at each stage.
    \item Analysis of the technical and organizational barriers that impede transitions between generations, with specific attention to the trust, knowledge, and cultural dimensions.
    \item A set of readiness indicators and metrics that signal when an organization is prepared to advance to the next generation of operational maturity.
    \item Production experience and quantitative results from navigating the full evolution from Generation 1 (manual) through Generation 5 (autonomous) operations at hyperscale.
    \item Identification of common failure modes and anti-patterns observed during operational evolution.
\end{enumerate}

This paper is complemented by a companion work~\cite{malik2025autonomous} that presents the detailed agentic AI architecture enabling Generation 5 autonomous operations, including the multi-agent coordination framework and safety mechanisms.

\subsection{Paper Organization}

Section~\ref{sec:maturity} presents the five-generation maturity model. Section~\ref{sec:transitions} analyzes the transitions between generations. Section~\ref{sec:barriers} examines barriers to evolution. Section~\ref{sec:metrics} proposes readiness indicators. Section~\ref{sec:production} presents production experience. Section~\ref{sec:lessons} discusses lessons learned. Section~\ref{sec:conclusion} concludes.

\section{Operational Maturity Model}
\label{sec:maturity}

We identify five distinct generations of operational maturity in cloud network infrastructure, each characterized by its architectural pattern, human role, AI role, and operational authority model.

\begin{table*}[t]
\centering
\caption{Five Generations of Operational Maturity in Cloud Network Infrastructure}
\label{tab:generations}
\begin{tabular}{@{}lllll@{}}
\toprule
\textbf{Generation} & \textbf{Pattern} & \textbf{Human Role} & \textbf{AI Role} & \textbf{Authority} \\
\midrule
Gen 1: Manual & SSH and CLI & Investigator, executor & None & Human-only \\
Gen 2: Scripted & Runbook automation & Triggerer, supervisor & None & Human-triggered \\
Gen 3: Rule-Based & Event-condition-action & Exception handler & Pattern matching & System-triggered, bounded \\
Gen 4: AI-Assisted & Recommendation engine & Decision maker & Advisor, analyst & Human-approved \\
Gen 5: Autonomous & Multi-agent orchestration & Auditor, policy setter & Perceiver, reasoner, actor & AI-driven, safety-bounded \\
\bottomrule
\end{tabular}
\end{table*}

\subsection{Generation 1: Manual Operations}

In the first generation, operational work is entirely human-driven. Engineers connect directly to network devices via SSH or console, execute diagnostic commands, interpret output through individual expertise, and apply remediations manually. Knowledge resides in the heads of experienced engineers and in informal documentation (wikis, chat logs, tribal knowledge).

\textbf{Characteristics:}
\begin{itemize}
    \item Mean time to resolution (MTTR) measured in hours to days
    \item Incident quality highly dependent on individual engineer skill
    \item Knowledge transfer occurs primarily through apprenticeship
    \item No systematic recording of diagnostic reasoning or remediation rationale
    \item Scalability limited by engineering headcount
\end{itemize}

\textbf{Failure mode:} Key-person dependency. When senior engineers leave, critical operational knowledge is lost permanently.

\subsection{Generation 2: Scripted Automation}

The second generation codifies common operational procedures into scripts and runbooks. Engineers still decide when and what to execute, but the execution itself is automated. This reduces human error in the remediation phase and enables junior engineers to handle incidents previously requiring senior expertise.

\textbf{Characteristics:}
\begin{itemize}
    \item MTTR reduced for known failure modes (minutes to hours)
    \item Scripts encode the "how" but not the "when" or "why"
    \item Runbook libraries grow organically, often without systematic testing
    \item Drift between documented procedures and actual scripts
    \item Still requires human diagnosis to select the appropriate script
\end{itemize}

\textbf{Failure mode:} Script sprawl. Organizations accumulate thousands of scripts with unclear ownership, overlapping functionality, and insufficient testing, making the automation layer itself a source of operational risk.

\subsection{Generation 3: Rule-Based Automation}

The third generation introduces event-driven triggering where systems autonomously execute predefined actions in response to specific conditions. Human engineers handle exceptions that fall outside the rule set.

\textbf{Characteristics:}
\begin{itemize}
    \item Autonomous handling of well-characterized failure modes
    \item Rules encode the "when" (conditions) and "what" (actions) but not reasoning
    \item Coverage limited to explicitly anticipated scenarios
    \item Boolean logic cannot represent uncertainty or partial matches
    \item Rule maintenance cost grows quadratically with complexity
\end{itemize}

\textbf{Failure mode:} Brittleness. Rule-based systems fail silently on novel failure modes that fall outside the predefined condition space, leading to missed incidents or incorrect actions on edge cases.

\subsection{Generation 4: AI-Assisted Operations}

The fourth generation introduces machine learning and AI to provide intelligent analysis and recommendations. AI systems analyze telemetry, correlate alerts, identify anomalies, and suggest remediations, but humans retain decision authority and execution control.

\textbf{Characteristics:}
\begin{itemize}
    \item AI handles pattern recognition across high-dimensional data
    \item Recommendations reduce cognitive load on engineers
    \item Humans retain full decision and execution authority
    \item AI errors are caught before execution by human review
    \item Slower than rule-based for known cases (human approval latency)
\end{itemize}

\textbf{Failure mode:} Recommendation fatigue. When AI systems generate too many recommendations or too many false positives, engineers begin ignoring them, negating the value of the AI system entirely.

\subsection{Generation 5: Autonomous Operations}

The fifth generation grants AI agents the authority to perceive, reason, decide, and act without requiring human approval for each action. Safety is maintained not through human gatekeeping but through architectural constraints: bounded authority, blast-radius containment, progressive trust, and automated rollback.

\textbf{Characteristics:}
\begin{itemize}
    \item End-to-end autonomous resolution for qualified incident categories
    \item MTTR measured in seconds to minutes
    \item Safety maintained through architectural constraints, not human oversight
    \item Human role shifts from operator to auditor and policy designer
    \item System handles novel scenarios through compositional reasoning
\end{itemize}

\textbf{Failure mode:} Authority creep. Without proper boundaries, autonomous systems gradually take actions beyond their validated scope, potentially causing cascading failures in areas where their reasoning has not been verified.

\section{Generational Transitions}
\label{sec:transitions}

The transitions between generations are not incremental improvements but architectural discontinuities that require fundamental changes in tooling, processes, and organizational structure.

\subsection{Transition 1 to 2: Codifying Knowledge}

The first major transition requires converting implicit, human-held operational knowledge into explicit, machine-executable scripts. This is primarily a knowledge management challenge.

\textbf{Technical requirements:}
\begin{itemize}
    \item Standardized device interaction abstractions (APIs replacing SSH)
    \item Version-controlled script repositories with testing frameworks
    \item Inventory systems mapping devices to appropriate procedures
\end{itemize}

\textbf{Organizational requirements:}
\begin{itemize}
    \item Cultural shift from "heroic individual" to "systematic process"
    \item Incentive structures that reward documentation and automation
    \item Dedicated time allocation for automation development
\end{itemize}

\subsection{Transition 2 to 3: Adding Decision Logic}

The second transition moves decision-making from humans to systems. The key challenge is defining the condition space precisely enough that automated actions are safe.

\textbf{Technical requirements:}
\begin{itemize}
    \item Reliable event streaming and processing infrastructure
    \item Formal specification of trigger conditions and guard clauses
    \item Monitoring and alerting for the automation system itself
    \item Rollback mechanisms for automated actions
\end{itemize}

\textbf{Organizational requirements:}
\begin{itemize}
    \item Acceptance that automated systems will occasionally take incorrect actions
    \item Incident review processes that evaluate automation failures
    \item Clear ownership boundaries between automated and manual domains
\end{itemize}

\subsection{Transition 3 to 4: Introducing Learning}

The third transition introduces AI systems that learn from data rather than operating on manually specified rules. This requires investing in data infrastructure and model development capabilities.

\textbf{Technical requirements:}
\begin{itemize}
    \item Comprehensive telemetry collection and storage
    \item Feature engineering pipelines for operational data
    \item Model training, validation, and deployment infrastructure
    \item Feedback loops connecting outcomes to model improvement
\end{itemize}

\textbf{Organizational requirements:}
\begin{itemize}
    \item Data science expertise embedded in operations teams
    \item Tolerance for probabilistic (non-deterministic) recommendations
    \item Processes for evaluating and calibrating AI system accuracy
\end{itemize}

\subsection{Transition 4 to 5: Granting Authority}

The most consequential transition grants AI systems authority to act independently. This is fundamentally a trust problem that cannot be solved purely through technical means.

\textbf{Technical requirements:}
\begin{itemize}
    \item Multi-agent orchestration with specialized capabilities
    \item Progressive autonomy framework with granular authority levels
    \item Safety mechanisms: blast-radius limits, rate limits, rollback
    \item Closed-loop verification confirming remediation success
    \item Comprehensive audit logging for accountability
\end{itemize}

\textbf{Organizational requirements:}
\begin{itemize}
    \item Trust framework defining conditions for authority delegation
    \item Governance structures for reviewing and adjusting AI authority
    \item Retraining programs for engineers transitioning to oversight roles
    \item Incident response procedures for AI-caused failures
\end{itemize}

\section{Barriers to Evolution}
\label{sec:barriers}

\begin{figure}[t]
\centering
\includegraphics[width=\columnwidth]{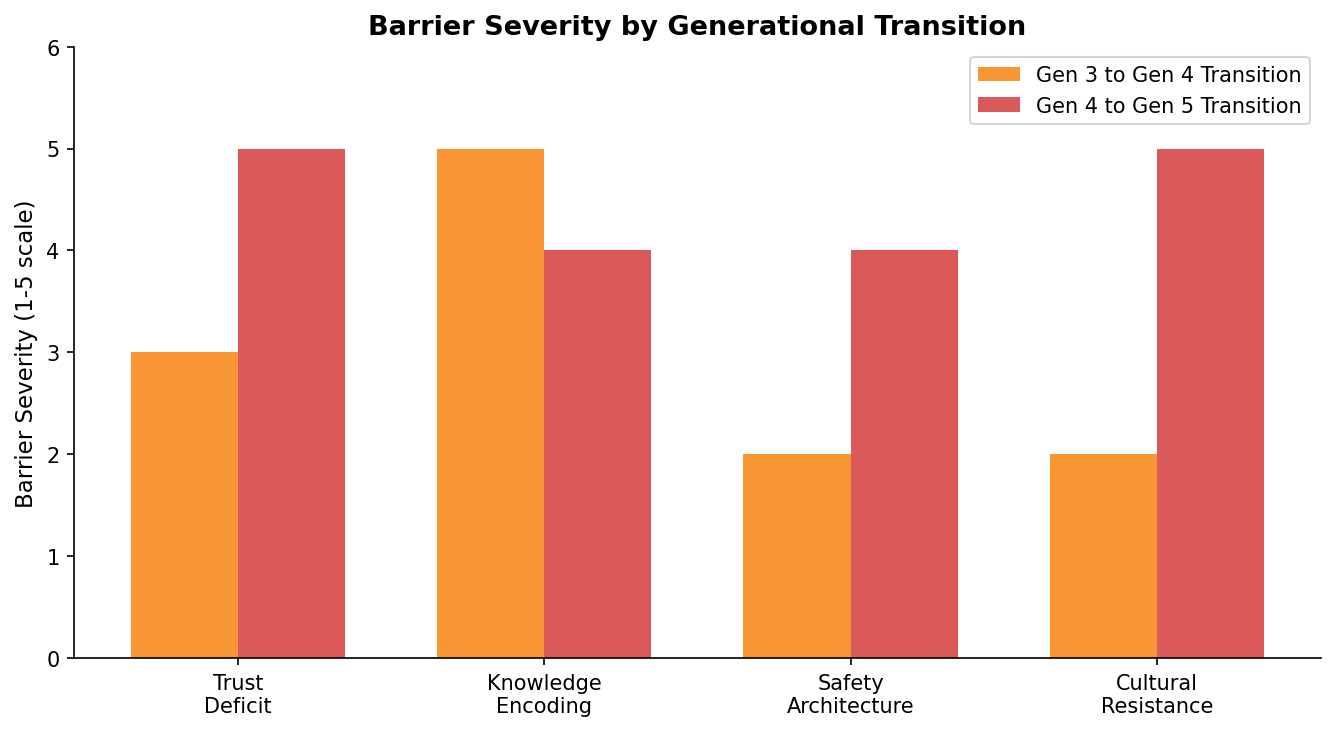}
\caption{Relative difficulty of barriers to operational evolution, based on production experience. Trust deficit consistently represents the hardest barrier to overcome.}
\label{fig:barriers}
\end{figure}

Our experience identifies four primary categories of barriers that impede progression through the maturity model (Figure~\ref{fig:barriers}).

\subsection{Trust Deficit}

The most persistent barrier is organizational trust. Even when AI systems demonstrate statistically superior performance to human operators, organizations resist granting operational authority due to perceived risk. This trust deficit manifests in several ways:

\begin{itemize}
    \item \textbf{Asymmetric accountability:} Human errors are treated as learning opportunities while AI errors are treated as system failures, creating an uneven evaluation standard.
    \item \textbf{Opacity concern:} Decision-makers cannot inspect AI reasoning with the same tools they use to evaluate human reasoning, creating discomfort with delegation.
    \item \textbf{Worst-case fixation:} Organizations focus on the worst possible AI failure rather than expected value, creating an impossibly high bar for adoption.
\end{itemize}

\textbf{Mitigation:} Progressive autonomy with measurable trust accumulation. Start with low-risk, high-frequency incidents where AI performance can be demonstrated and trust built incrementally.

\subsection{Knowledge Encoding}

Operational knowledge in most organizations exists in forms that are inaccessible to AI systems: in engineers' heads, in unstructured wiki pages, in chat logs, and in code comments. Converting this knowledge into structured, machine-executable form requires sustained effort.

\begin{itemize}
    \item \textbf{Tacit knowledge:} Experienced engineers often cannot articulate their diagnostic reasoning explicitly.
    \item \textbf{Context dependence:} The correct action depends on contextual factors (time of day, current load, recent changes) that are rarely documented.
    \item \textbf{Knowledge decay:} Infrastructure changes faster than documentation, creating persistent drift.
\end{itemize}

\textbf{Mitigation:} Structured knowledge encoding through observed behavior. Rather than asking engineers to document their reasoning, capture their actions and decisions in structured form and use these records to build executable knowledge bases.

\subsection{Safety Architecture}

Designing safety mechanisms that enable rather than prevent autonomy is an unsolved challenge for many organizations. Common failure patterns include:

\begin{itemize}
    \item \textbf{Over-constraining:} Safety rules that prevent the AI system from taking any meaningful action, rendering autonomy meaningless.
    \item \textbf{Under-constraining:} Insufficient boundaries that allow the AI system to take dangerous actions in edge cases.
    \item \textbf{Static boundaries:} Fixed safety limits that do not adapt to demonstrated competence or changing conditions.
\end{itemize}

\textbf{Mitigation:} Layered safety with dynamic boundaries. Implement multiple independent safety layers (rate limits, blast-radius caps, verification steps, rollback triggers) and adjust boundaries based on demonstrated performance.

\subsection{Cultural Resistance}

Autonomous AI systems change the fundamental nature of operational work. Engineers who built their careers on hands-on troubleshooting may resist systems that diminish the value of their hard-won expertise.

\begin{itemize}
    \item \textbf{Identity threat:} Engineers whose professional identity centers on troubleshooting skill feel threatened by automation of their core competency.
    \item \textbf{Skill atrophy concern:} If AI handles routine incidents, will engineers lose the ability to handle complex ones?
    \item \textbf{Career path uncertainty:} Unclear progression for engineers in organizations with autonomous operations.
\end{itemize}

\textbf{Mitigation:} Reframe the engineering role as "teaching and governing AI systems" rather than "being replaced by AI systems." Create career paths that value knowledge encoding, safety design, and AI governance.

\section{Readiness Indicators}
\label{sec:metrics}

We propose a set of indicators that signal organizational readiness to advance between generations. These indicators span technical infrastructure, operational processes, and organizational culture.

\begin{table*}[t]
\centering
\caption{Readiness Indicators for Generational Transitions}
\label{tab:readiness}
\begin{tabular}{@{}llp{7cm}@{}}
\toprule
\textbf{Transition} & \textbf{Category} & \textbf{Indicator} \\
\midrule
1 $\rightarrow$ 2 & Technical & API coverage $>$80\% of device operations \\
 & Process & Runbook documentation exists for $>$60\% of incident types \\
 & Cultural & Engineers allocated $>$20\% time to automation \\
\midrule
2 $\rightarrow$ 3 & Technical & Event streaming with $<$30s end-to-end latency \\
 & Process & Automated testing for $>$70\% of operational scripts \\
 & Cultural & Post-incident reviews evaluate automation opportunities \\
\midrule
3 $\rightarrow$ 4 & Technical & $>$6 months of structured telemetry and outcome data \\
 & Process & Model evaluation framework with defined accuracy thresholds \\
 & Cultural & Operations team includes or partners with data science \\
\midrule
4 $\rightarrow$ 5 & Technical & Multi-agent framework with independent safety layers \\
 & Process & Progressive authority model with measurable trust accumulation \\
 & Cultural & Leadership endorsement of AI operational authority \\
\bottomrule
\end{tabular}
\end{table*}

\subsection{Technical Readiness}

Technical readiness encompasses the infrastructure, tooling, and data foundations required for each generation:

\begin{itemize}
    \item \textbf{API coverage:} Percentage of operational actions available via programmatic interfaces (vs. CLI-only).
    \item \textbf{Telemetry completeness:} Percentage of operational state observable through automated collection.
    \item \textbf{Action reversibility:} Percentage of automated actions that can be rolled back programmatically.
    \item \textbf{Verification coverage:} Percentage of remediations with automated success/failure verification.
\end{itemize}

\subsection{Process Readiness}

Process readiness measures the maturity of operational procedures and governance:

\begin{itemize}
    \item \textbf{Knowledge formalization:} Ratio of documented vs. tribal knowledge for operational procedures.
    \item \textbf{Automation testing:} Coverage and recency of automated tests for operational tooling.
    \item \textbf{Outcome tracking:} Percentage of incidents with structured resolution data suitable for learning.
    \item \textbf{Authority governance:} Existence and enforcement of policies governing AI operational authority.
\end{itemize}

\subsection{Cultural Readiness}

Cultural readiness captures organizational attitudes and behaviors:

\begin{itemize}
    \item \textbf{Automation investment:} Percentage of engineering time dedicated to building vs. using tools.
    \item \textbf{Error tolerance:} Organizational response to automation failures (learning vs. blame).
    \item \textbf{Role evolution:} Clarity of career paths for engineers in increasingly automated environments.
    \item \textbf{Leadership support:} Active sponsorship of autonomy initiatives by engineering leadership.
\end{itemize}

\begin{figure}[t]
\centering
\includegraphics[width=\columnwidth]{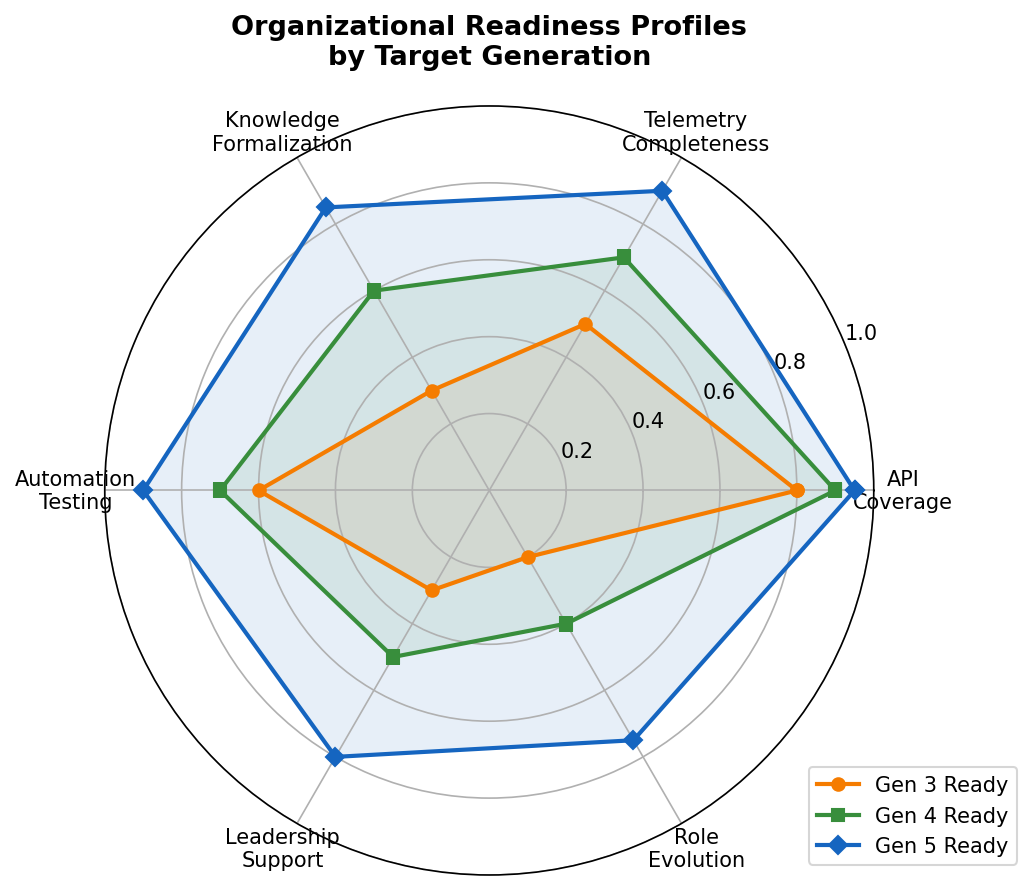}
\caption{Organizational readiness radar across three dimensions: technical infrastructure, process maturity, and cultural preparedness. Balanced advancement across all dimensions is critical for sustainable evolution.}
\label{fig:readiness}
\end{figure}

\section{Production Experience}
\label{sec:production}

We present quantitative results from navigating the full maturity evolution in a hyperscale cloud network operations environment over approximately 18 months.

\subsection{Starting Conditions}

The environment at the start of our evolution effort exhibited characteristics typical of a mature Generation 2 organization with some Generation 3 elements:

\begin{itemize}
    \item Over 12 million managed network devices
    \item Approximately 3,000 operational incidents per day
    \item Mean time to resolution of 10.2 hours for common incident categories
    \item 847 operational runbooks, of which approximately 60\% were partially automated
    \item Incident resolution quality highly variable across engineering shifts
\end{itemize}

\subsection{Evolution Timeline}

\begin{table}[h]
\centering
\caption{Timeline of Operational Maturity Evolution}
\label{tab:timeline}
\begin{tabular}{@{}lll@{}}
\toprule
\textbf{Phase} & \textbf{Duration} & \textbf{Focus} \\
\midrule
Gen 3 consolidation & 3 months & Rule standardization \\
Gen 4 introduction & 4 months & AI recommendation engine \\
Gen 4 to 5 transition & 5 months & Progressive authority \\
Gen 5 expansion & 6 months & Coverage growth \\
\bottomrule
\end{tabular}
\end{table}

\subsection{Quantitative Results}

The evolution produced measurable improvements across all key operational metrics (Figure~\ref{fig:mttr}):

\begin{figure}[t]
\centering
\includegraphics[width=\columnwidth]{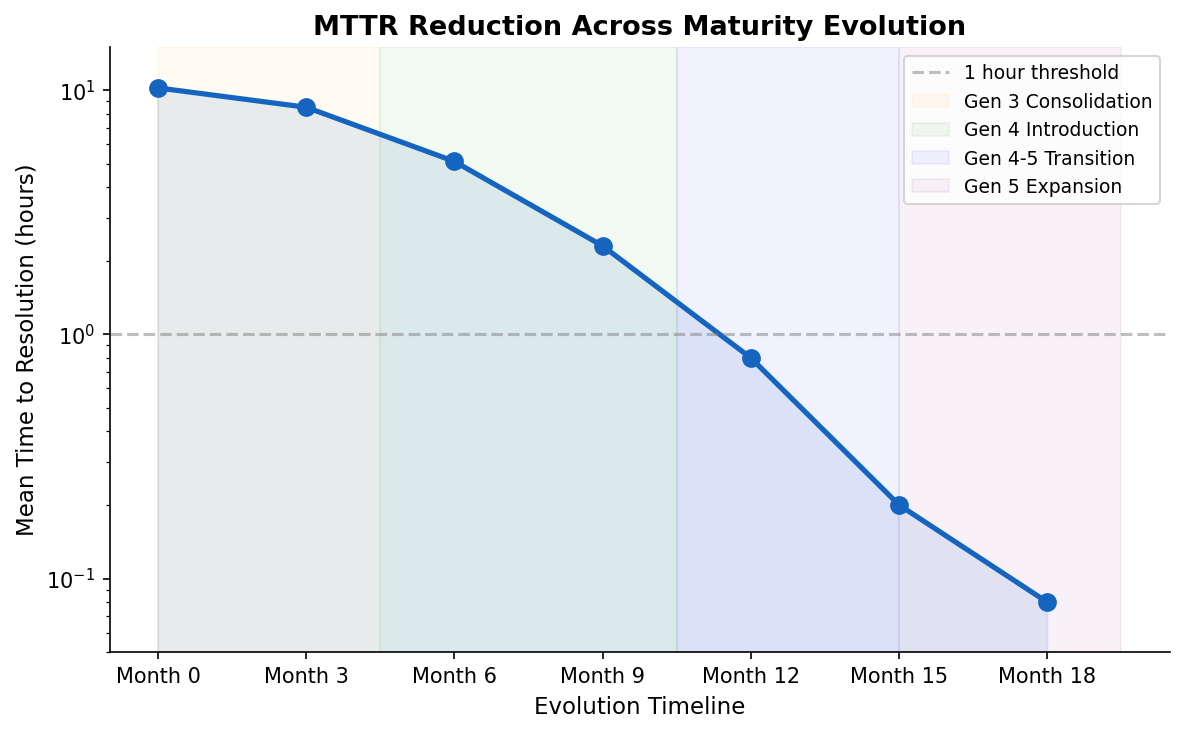}
\caption{Mean time to resolution (MTTR) across operational generations. Gen 5 autonomous operations achieve 122x improvement over the Gen 2/3 baseline.}
\label{fig:mttr}
\end{figure}

\begin{itemize}
    \item \textbf{MTTR reduction:} From 10.2 hours (Gen 2/3 baseline) to 5 minutes (Gen 5), representing a 122x improvement for autonomously resolved incidents.
    \item \textbf{Autonomous resolution rate:} 96.6\% of qualified incidents resolved without human intervention.
    \item \textbf{Coverage expansion:} From 12 incident categories at Gen 5 launch to 47 categories after 6 months of expansion.
    \item \textbf{Safety record:} Zero severity-1 or severity-2 incidents caused by autonomous actions over the full deployment period.
    \item \textbf{Engineer redeployment:} 40\% of on-call engineering time reclaimed for proactive improvement work.
\end{itemize}

\begin{figure}[t]
\centering
\includegraphics[width=\columnwidth]{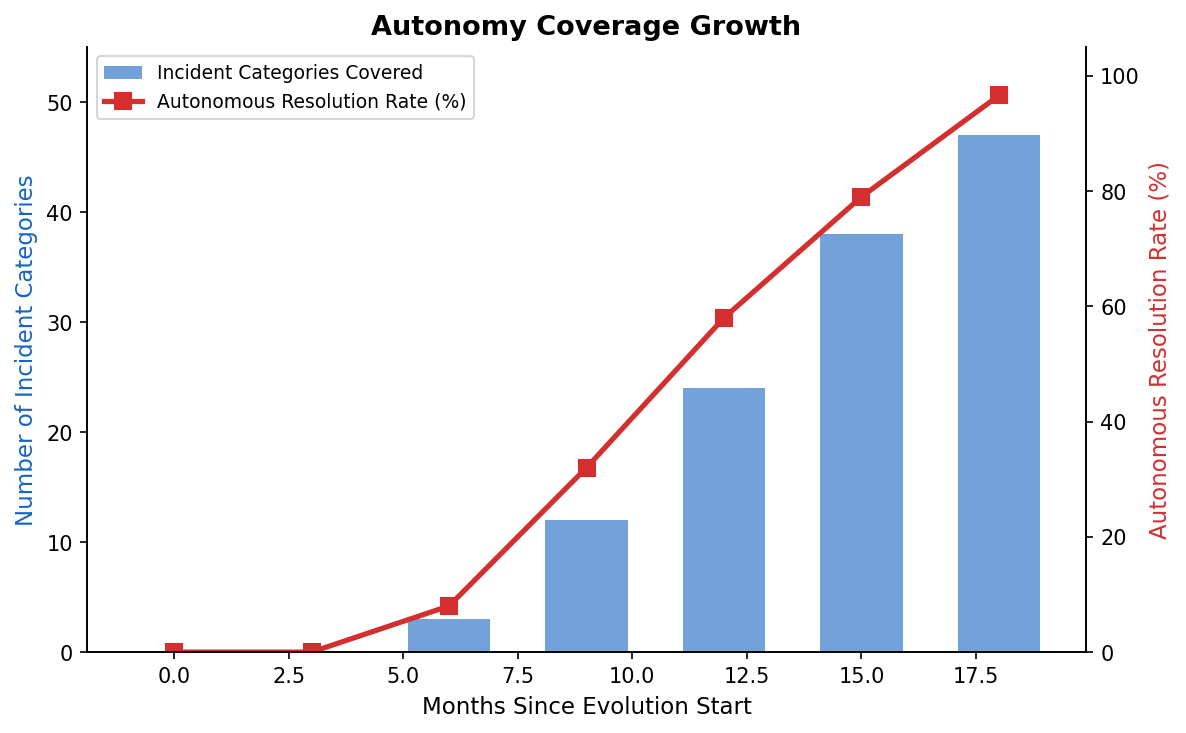}
\caption{Autonomous coverage expansion over the 6-month Gen 5 growth phase, showing incident categories handled autonomously growing from 12 to 47 categories.}
\label{fig:coverage}
\end{figure}

\subsection{Transition Challenges Encountered}

Several specific challenges emerged during the transition:

\textbf{Gen 3 to 4 transition:} The primary challenge was telemetry quality. Existing monitoring systems generated alerts optimized for human interpretation (natural language descriptions, context-dependent severity) rather than machine consumption (structured fields, consistent taxonomy). A 3-month effort to restructure telemetry was required before AI models could be trained effectively.

\textbf{Gen 4 to 5 transition:} The primary challenge was trust building. Despite the AI system demonstrating 94\% recommendation accuracy during the Gen 4 phase, obtaining organizational approval for autonomous execution required extensive demonstration, incremental authority expansion, and detailed safety analysis.

\section{Lessons Learned}
\label{sec:lessons}

\subsection{Start with the Boring Problems}

The most effective path to autonomous operations begins with high-frequency, low-complexity incidents rather than challenging edge cases. These incidents offer:

\begin{itemize}
    \item Sufficient training data for AI systems
    \item Low consequence of errors during early trust-building
    \item High visibility of operational improvement (measurable MTTR reduction)
    \item Rapid iteration cycles due to frequent occurrence
\end{itemize}

Attempting to automate complex, rare incidents first is a common anti-pattern that produces impressive demos but limited production value.

\subsection{Trust is Earned in Milliseconds but Lost in Seconds}

Building organizational trust in autonomous systems requires extended periods of demonstrated reliability. A single visible failure can eliminate months of accumulated trust. This asymmetry has architectural implications:

\begin{itemize}
    \item Safety mechanisms must be conservative during trust-building phases
    \item Failures must be contained, visible, and automatically remediated
    \item Success must be continuously measured and communicated
    \item Authority expansion must be gradual and reversible
\end{itemize}

\subsection{Knowledge Encoding is the Hardest Problem}

Neither pure machine learning nor pure knowledge engineering alone suffices for operational AI. The most effective approach combines:

\begin{itemize}
    \item Structured encoding of well-understood procedures (deterministic)
    \item Machine learning for pattern recognition and anomaly detection (probabilistic)
    \item Large language models for reasoning about novel combinations (compositional)
    \item Human oversight for truly unprecedented situations (exceptional)
\end{itemize}

\subsection{Safety Enables Rather Than Prevents Autonomy}

Counter-intuitively, investing heavily in safety mechanisms accelerates rather than impedes the adoption of autonomous operations. Strong safety guarantees reduce organizational risk perception, enabling faster authority expansion. Organizations that skimp on safety infrastructure find themselves permanently stuck at Generation 4 because they cannot build sufficient trust for the transition to Generation 5.

\subsection{Culture Eats Architecture}

The most technically sophisticated autonomous operations system will fail in an organization that does not support it culturally. Successful adoption requires:

\begin{itemize}
    \item Engineering leadership that actively champions AI authority
    \item Career paths that reward teaching AI rather than heroic troubleshooting
    \item Incentive structures aligned with automation rather than ticket volume
    \item Psychological safety for engineers adapting to new roles
\end{itemize}

\section{Related Work}

The AIOps landscape has been surveyed extensively~\cite{notaro2021survey, dang2019aiops}. Our work differs from these surveys by providing a practitioner-focused maturity model grounded in production deployment rather than a taxonomy of research approaches.

Operational maturity models exist for DevOps~\cite{forsgren2018accelerate} and SRE~\cite{beyer2016site} practices, but these do not address the specific challenges of AI-driven operational authority. Our model extends this lineage by adding the trust and authority dimensions unique to autonomous AI systems.

The progression from human-in-the-loop to human-on-the-loop to human-off-the-loop has been explored in autonomous vehicle~\cite{sae2021j3016} and military systems~\cite{scharre2018army} contexts. We adapt these concepts to the operational domain, where the consequences and timescales differ significantly from physical-world autonomy.

Recent work on LLM-based agents for operations~\cite{chen2024autocoderover, zhang2024autocoderover} has demonstrated the potential of language model agents for software engineering tasks. Our work extends these ideas to network infrastructure operations, where the safety requirements and scale challenges are substantially different.

\section{Conclusion}
\label{sec:conclusion}

The evolution from reactive to autonomous operations is not a single technology adoption but a multi-generational journey requiring co-evolution of architecture, knowledge management, trust frameworks, and organizational culture. We have presented a five-generation maturity model that provides a structured framework for understanding this evolution, identified the barriers and readiness indicators for each transition, and documented production experience navigating the full arc at hyperscale.

Our key finding is that the transition from AI-assisted (Generation 4) to autonomous (Generation 5) operations represents the most significant discontinuity in the maturity model. It is not merely a technical challenge of building better AI but a sociotechnical challenge of building trust, encoding knowledge, designing safety, and evolving culture simultaneously.

Organizations seeking to adopt autonomous operations should expect an 18-to-24-month journey, invest disproportionately in safety infrastructure and knowledge encoding, begin with high-frequency low-complexity incidents, and treat trust as their most valuable and fragile asset.

The future of network operations is not a choice between human and AI operators but an evolving partnership where the boundaries of authority shift progressively based on demonstrated competence, safety guarantees, and organizational readiness.

\section*{Acknowledgments}

This work reflects the collective efforts of the Azure Networking operations organization. The author thanks team members who contributed to developing, deploying, and validating these systems at production scale.

\bibliographystyle{IEEEtran}
\bibliography{references}

\end{document}